\documentclass[aps,prl,showpacs,floats,twocolumn,floats,superscriptaddress,floatfix]{revtex4-1}
\usepackage{graphicx}
\usepackage{bm}
\usepackage{amsfonts}
\usepackage{color}
\usepackage{amsmath}    % need for subequations
\usepackage{epsfig}
\newcommand{\K}{\mathrm{K}}
\newcommand{\A}{\mathrm{A}}
\newcommand{\ictsaddress}{International Centre for
  Theoretical Sciences, Tata Institute of Fundamental Research,
  Bangalore 560089, India}
\newcommand{\iiscaddressCSA}{Department of Computer Science and Automation, Indian 
Institute of Science, Bangalore 560012, India}

\newcommand{\sppuaddress}{Department of Physics, Savitribai Phule Pune
University, Pune 411007, India}
\newcommand{\uiucaddress}{John A. Paulson School of Engineering and Applied Sciences, Harvard University, Cambridge, MA 02138}
\begin{document}

\title{Flocking of Active Particles in a Turbulent Flow}
\author{Anupam Gupta}\email{anupam1509@gmail.com}
\affiliation{\uiucaddress}
\author{Amal Roy}
\email{amalroy@iisc.ac.in}
\affiliation {\iiscaddressCSA}
\author{Arnab Saha}
\email{sahaarn@gmail.com}
\affiliation{\sppuaddress}
\author{Samriddhi Sankar Ray}
\email{samriddhisankarray@gmail.com}
\affiliation{\ictsaddress}
\date{\today}% It is always \today, today,
             %  but any date may be explicitly specified


\begin{abstract}

	We investigate the effect of cooperative interactions in an
        ensemble of microorganisms, modelled as
	self-propelled disk-like  and rod-like particles, in a three-dimensional
	turbulent flow to show flocking as an emergent phenomenon. Building on
	the work by Choudhary, \textit{et al.} [Europhys.
	Lett. \textbf{112}, 24005 (2015)] for two-dimensional systems, and combining ideas from active
	matter  and turbulent transport, we show that non-trivial correlations
	between the flow and individual dynamics are \textit{essential} for the
	microorganisms to flock in, for example, a turbulent three-dimensional, marine
	environment. Our results may have implications especially in the
	modelling of artificial microswimmers in a \textit{hostile}
	environment.

\end{abstract}
\maketitle

The understanding of collective, cooperative motion of organisms is
one of the most important problems in areas spanning physical
biology, soft matter, and statistical physics.
This is largely due to its ubiquity in the
natural world spanning a range of scales from macroscopic (birds and fish) to
microscopic (bacteria and plankton) organisms~\cite{Parrish}. Therefore it is
reasonable to expect that  insights to this phenomenon is critically related to
understanding self-organised behaviour as well as ecological and evolutionary
strategies.  Such insights, drawn from a variety of models~\cite{Topaz,Eftimie,Giardina, Alaimo, Dunkel}, show 
that combining long-range attraction, short-range repulsion and alignment rules
between self-propelling individuals of model active systems lead to self-organised
collective dynamics of the individuals.  It thus allows us qualitative as
well as quantitative understanding of the dynamics of real active
systems such as cells~\cite{Elod,Chen}, autophoretic colloids~\cite{Buttinoni}, 
macroscopic, yet small, organisms~\cite{Buhl,Couzin,Kelley} or droplets~\cite{Thutupalli}, 
synchronised oscillators~\cite{strogatz} and bacterial lattices~\cite{lattice}.  

Developing an understanding of flocking has been largely confined to the
problem of directed motion where the effects of the ambient medium, air or
water---typically {\it noisy}, {\it random} and {\it spatio-temporally}
complex---and the shapes and sizes of the individuals, are ignored.  Such an
assumption for the natural world, where the environment can be non-trivial and
even turbulent (such as for marine life), is an over-simplification.  Indeed,
the few studies which have tried to answer this question use either model
flows~\cite{nickNJP,baggaley} with mass-less, point-sized organisms or the
relatively more realistic and instructive, but still academic, case of
two-dimensional turbulent flows with finite-sized {\it spherical}
particles~\cite{ashokEPL} and, more recently, flocking colloids in a randomly
and artificially disordered environment~\cite{disorder}.  These studies,
however, strongly suggest the need to study whether collective motion can exist
in the most general, even if simplified, setting.  Therefore in this paper, we
now show how and why flocking emerges---surprisingly---for microorganisms in a
turbulent environment by combining ideas from turbulent
transport~\cite{bec1,bec2,ray-PoF} and active
matter~\cite{vicsek,vicsek1,SriramRMP,HartmutRMP} and elucidate the critical
role of the non-trivial correlations between the flow and individual dynamics.
Our work shows that the principles of fluid mechanics leave little choice for
organisms but to flock based on their sizes and shapes. Thus the inevitable
ubiquity of this phenomenon.

For several reasons, the collective behaviour of self-propelled  (active)
individuals depends on their ambient medium.  Firstly, self propulsion of
active particles---which is key to their self-organisation---arises from the
fact that each individual is able to convert the available free energy into
directed motion~\cite{SriramRMP, HartmutRMP}. Examples include eukaryotic cells
which utilize the energy produced by the hydrolysis of available ATP to
self-organise and form tissue during morphogenesis~\cite{OnGrowthAndForm} or
Janus colloids with their asymmetric surface chemistry converting the
difference in the chemical potential across their surfaces and the surrounding
medium into self-propulsion~\cite{Chaikin2013, Paxton}. Secondly, a fluid
medium usually allows for long-ranged hydrodynamic interaction among the active
particles, which, in turn, influences their
self-organisation~\cite{Adhikari2016, Saha2016, Stark2014}. But these
interactions, which closely represents the hydrodynamics of various soft
systems such as colloidal suspensions~\cite{Mazur1982, MazurVansaarloos1982},
are at Reynolds numbers which are essentially zero.  Hydrodynamics at low
Reynolds number~\cite{Purcell1977} is, however, not applicable to the
collective dynamics of the self-propelled microorganisms in a turbulent
environment. 

For organisms which move and flock in air or water,  a na\"ive guess would be
that the flocks ought to break up in the presence of such strong
perturbations~\cite{nickNJP}.  However, observations show that a wide class of
organisms are able to overcome such strong perturbations and show evidence of
collective behaviour~\cite{stocker,durham}.  Therefore it is important 
to investigate if the active nature of
microorganisms and their strategy to self-organise can overcome their Stokesian drag and turbulent
mixing, to show organised, flocking behaviour in a three-dimensional
(turbulent) fluid environment? And, if so, are there preferred sizes and
shapes which lead to flocking as a spontaneously emergent phenomenon in a
collection of self-propelled particles in a turbulent flow?

\begin{figure}
\begin{center}
\includegraphics[width=1.0\columnwidth]{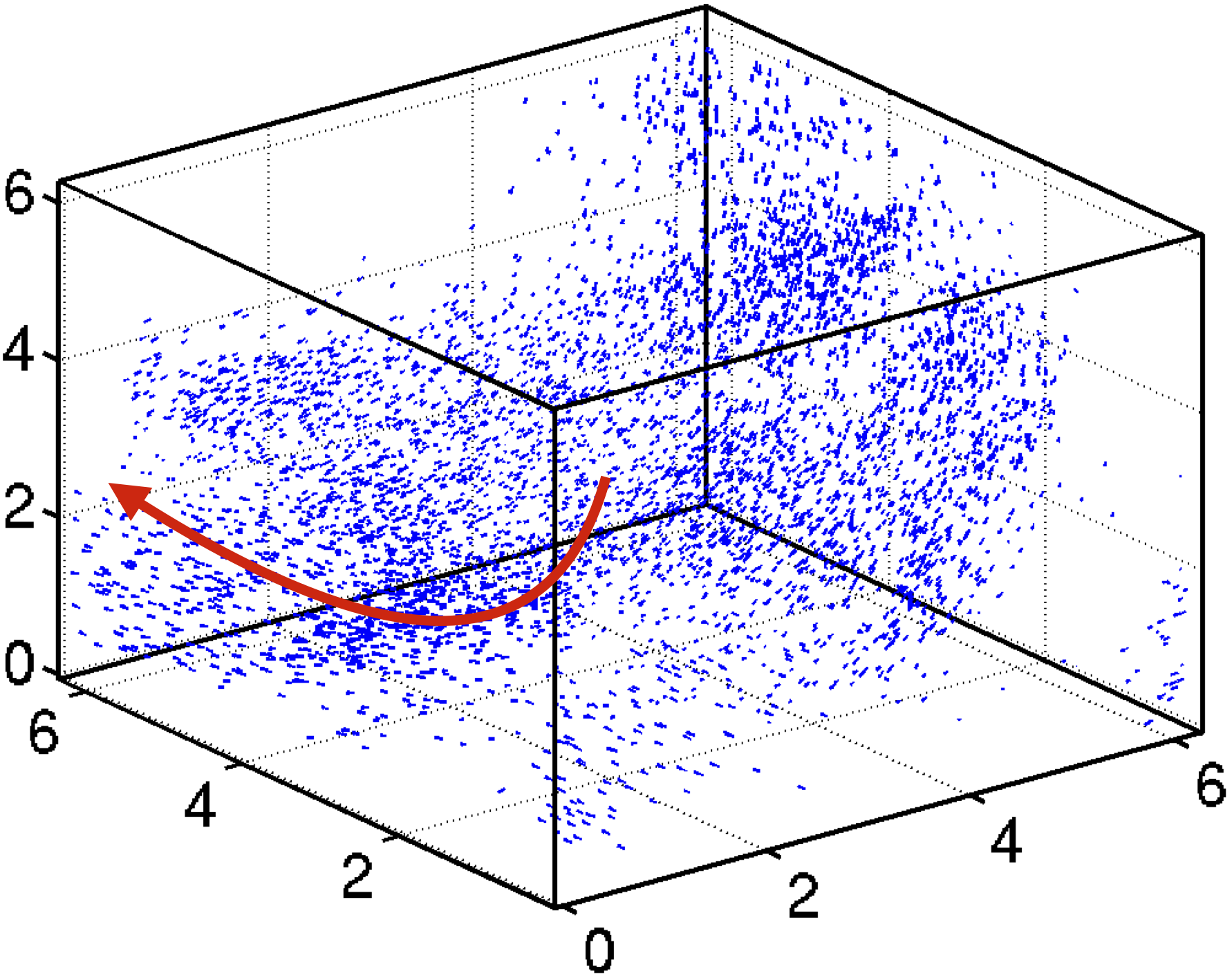}
\end{center}
\caption{A representative snapshot of a randomly chosen subset, for clarity, of a population of microorganisms 
which shows collective, flocking behaviour. The blue arrows are the individual velocity vectors; the thick red 
	arrow is a guide to the eye showing the average direction in which the flock is moving. (See also the online 
	videos from our simulations~\cite{video}.)}
\label{fig:snapshot}
\end{figure}

We therefore perform detailed numerical simulations of the motion of a
suspension of self-propelled active particles, of different shapes and sizes,
in three-dimensional turbulence. We use a standard pseudo-spectral method to 
obtain turbulent solutions of the driven, incompressible Navier-Stokes equation and 
choose parameters to ensure a Taylor-scale Reynolds number $Re_\lambda = 130$ which 
is consistent with typical marine settings~\cite{jimenez}.
These microorganisms, modelled as spheroids,  have both rotational and translational degrees of freedom
and their dynamics is governed by (a) the drag due to the carrier flow which
depends on their individual sizes and shapes, (b) their self-propelled velocity, and (c)
their short-ranged alignment interactions with other individuals
within a neighbourhood. We ensure that 
the particle sizes are smaller than the characteristic Kolmogorov length scales of the fluid.
The dynamics of such particles are characterised by their (unit)
orientation vector ${\bf \hat p}_i$ following~\cite{jeffery} 
\begin{equation}  
\frac{d{\bf \hat{p}}_i}{dt}={\bf{\Omega}}.{\bf \hat{p}}_i+\frac{\alpha^2-1}{\alpha^2+1}\left[{\bf{S. \hat p}}_i-({\bf \hat{p}}_i.{\bf {S}}.{\bf \hat{p}}_i){\bf \hat{p}}_i\right]
\label{p_update} 
\end{equation}
and their translational velocity vector, determined from the linear Stokes drag model (valid for small particles) for 
non-spherical particles~\cite{bec1,bec2,Maxey}
\begin{equation}
\frac{d{\bf r}_i}{dt} = {\bf v}_i; \quad \frac{d{\bf v}_i}{dt} =  -\frac{\A^{\mathrm T}\K \A}{6\pi a \alpha} \frac{[{\bf v}_i - {\bf u}({\bf r}_i,t)]}{\tau_p},
\label{eq}
\end{equation}
where $i$ is the particle index.  Not surprisingly the instantaneous orientation is determined by the local
geometry of the advecting flow $\bf u$, namely the symmetric ${\bf S}$ (strain
rate) and antisymmetric ${\bf \Omega}$ (vorticity) tensors of the
fluid-velocity-gradients at the particle positions and inertial effects in the
rotational dynamics are negligible as shown in previous
studies~\cite{anderson-prl,marchioli}.  Inertial effects are however important
in the translational degrees of freedom (Eq.~\eqref{eq}) where the effective
drag on the individual depends on the instantaneous orientation in the flow as
well as its shape $\alpha$, the ratio of the semi-major and minor axes of the
particles~\cite{mortensenPoF08,challabotlajfm15}: $\alpha = 1$ is a spherical;
$\alpha < 1$ is disk-like (oblate)  and $\alpha > 1$ is rod-like (prolate).
Thus, the resistance tensor $K$ and the orthogonal transformation matrix $A$
are necessary to take into account the shape of the
microorganism~\cite{mortensenPoF08,AmalPRE} and this leads to a time-dependent
drag coefficient, in contrast with the constant drag associated with a
spherical particle. However, for convenience we define the \textit{average} Stokes number $St =
\tau_p/\tau_\eta$~\cite{anderson-prl,AmalPRE} to quantify our results as well
as compare microorganisms of different shapes; $\tau_\eta$ is the
characteristic small-scale Kolmogorov time-scale of the ambient fluid. Our
particles are of course {\it active} with a self-propelled velocity along the
orientation vector $v_0{\bf \hat p}_i$. Coupling this with translational
equations of motion, the instantaneous velocity vector of the $i$-th individual
is a superposition of these two competing effects, namely ${\bf v}_i = {\bf
v}_i + v_0{\bf \hat p}_i.$

\begin{figure*}
\begin{center}
\includegraphics[width=0.325\textwidth]{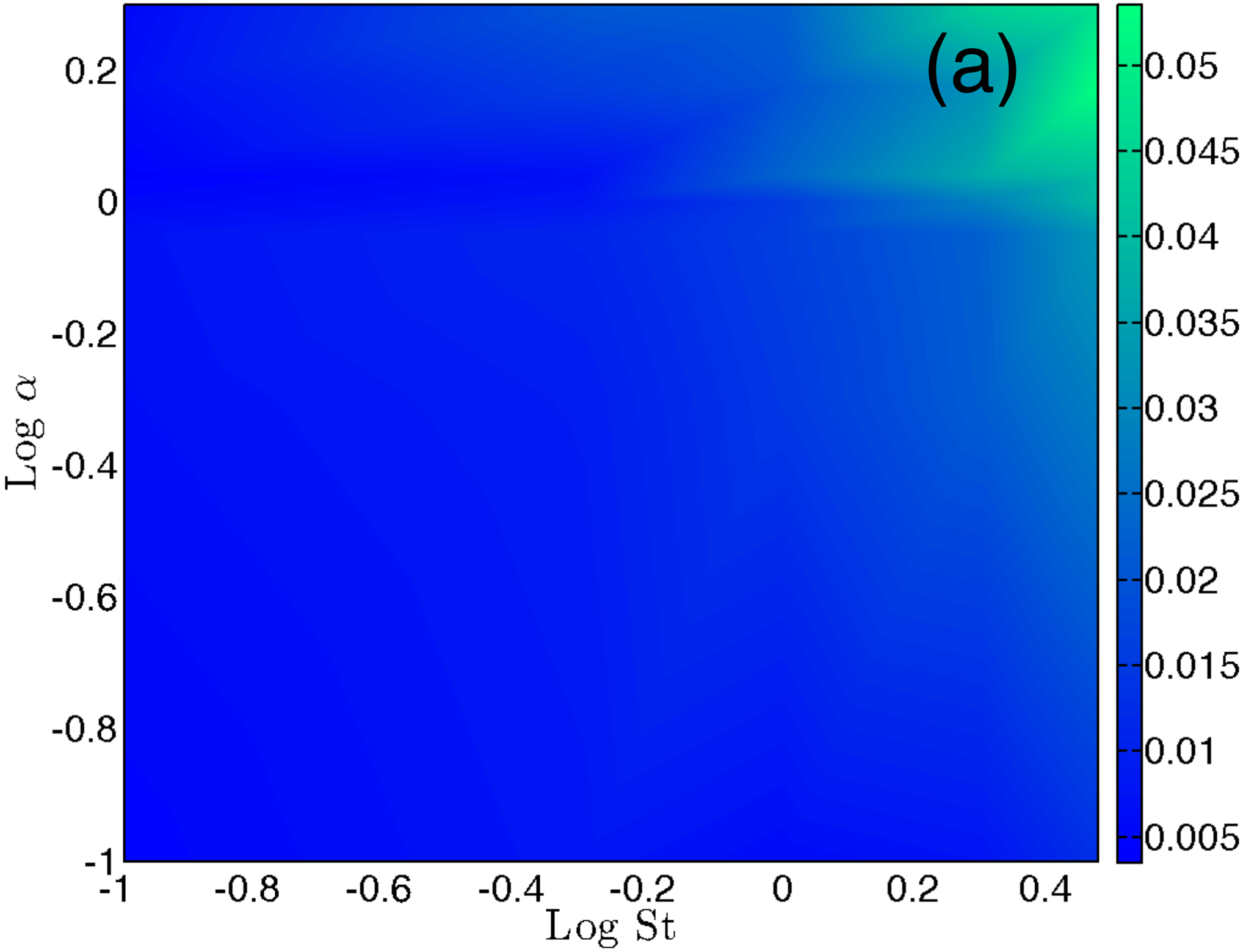}
\includegraphics[width=0.325\textwidth]{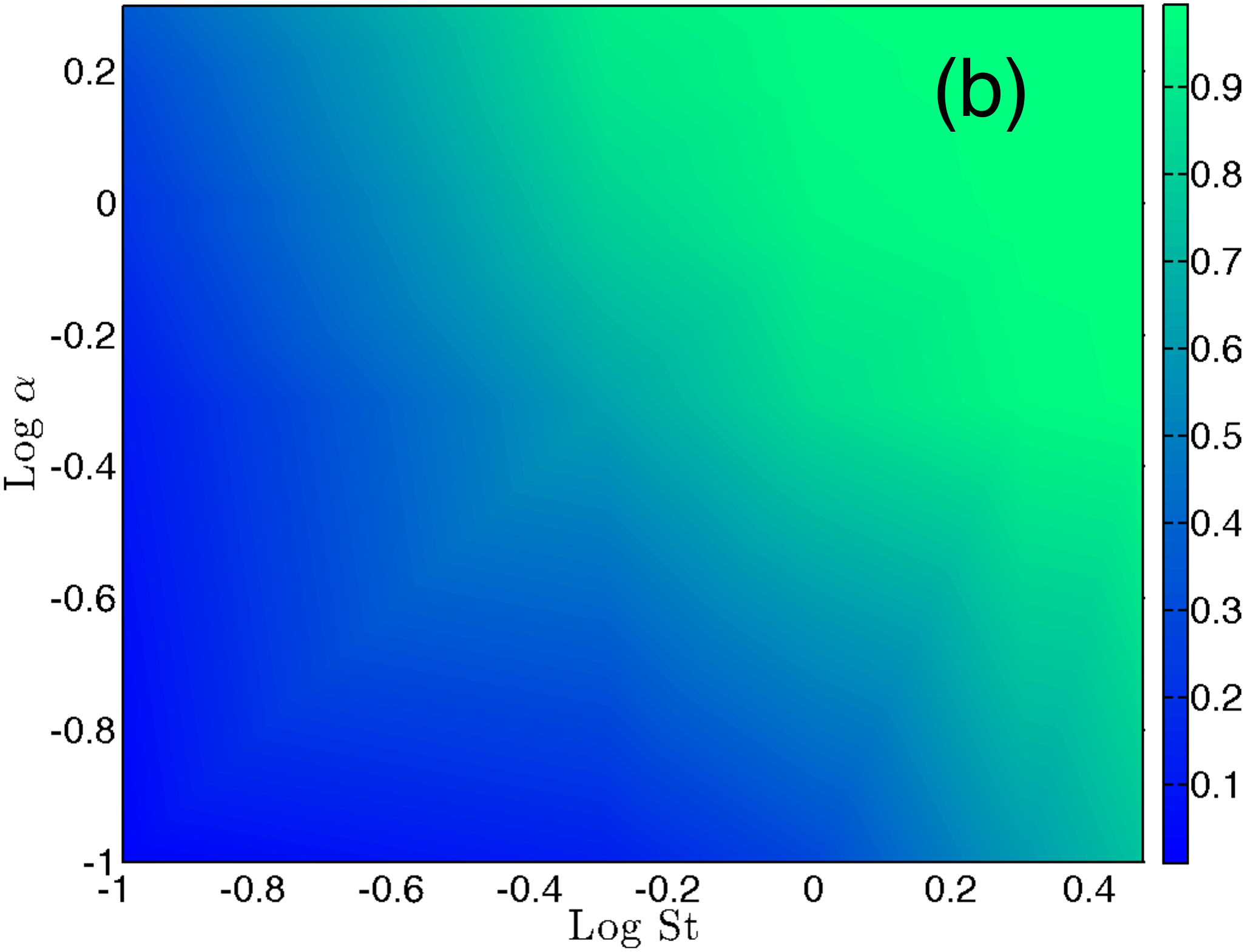}
\includegraphics[width=0.33\textwidth]{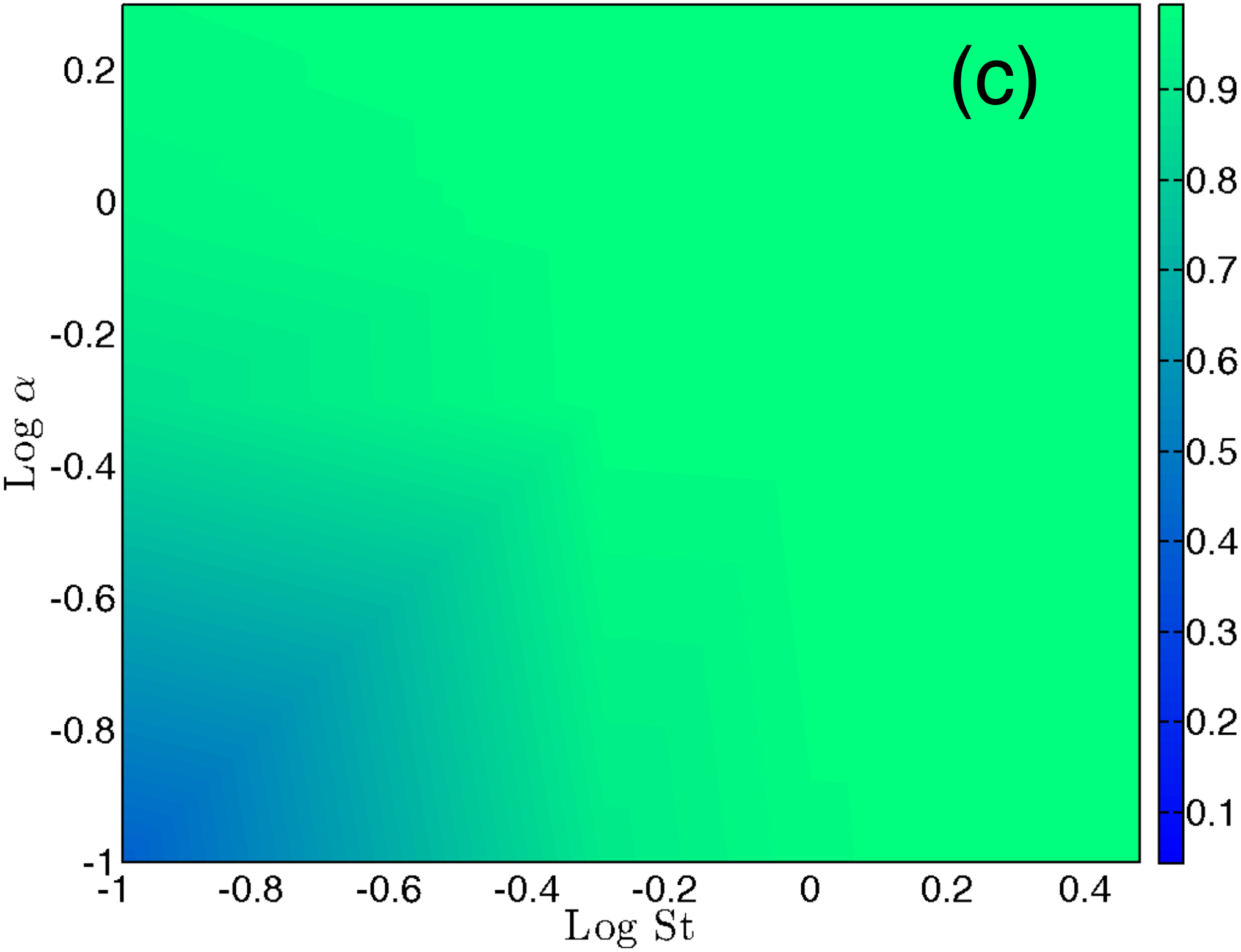}
\end{center}
\caption{Representative log-log pseudo-color plots of the global order parameter as a function 
of the size and shape for (a) $r_{\rm int} = 0.03\pi$ and $v_0 = 0$, (b) $r_{rm int} = 0.2\pi$ and $v_0 = 0.2u_\eta$, and  (c) $r_{\rm int} = 
0.4\pi$ and $v_0 = 1.7u_\eta$.}
\label{fig:global}
\end{figure*}

Let us finally introduce interactions in a collection of such active microorganisms. 
We use Vicsek-like interactions~\cite{vicsek}
to ensure that at discrete time intervals $\Delta T$ each individual $i$ orients itself 
along the average direction ${\bf \hat n}_i^{\rm avg}$ of all individuals $N_{\rm int}$ (with the same shape and size) 
within a radius of interaction $r_{\rm int}$ of it. This radius of interaction is a way to ensure, as is the case 
in actual living organisms, that an individual is not able to {\it see} others in the population 
who are physically far away. Of course this assessment of the average orientation cannot be perfect. To account for 
this imperfection we introduce an additive small random noise in the 
calculation of  ${\bf \hat n}_i^{\rm avg}$ for each individual. This implies 
\begin{equation} 
{\bf v}_i(t_\star)= |{\bf v}_i (t_\star)|{\bf \hat n}^{\rm avg}_i, 
\end{equation} 
where  $t_\star = m\Delta T$ ($m$ is an integer). A similar 
strategy is used for the re-orientation at discrete time intervals for the vector ${\bf \hat p}_i$. 
Between successive reorientations, particles evolve through the linear, Stokes drag model 
and its self-propelled velocity as described. We report results from 3 different values of 
$r_{\rm int}$, namely, $0.03\pi$, $0.2\pi$, and $0.4\pi$. 
(This strategy is also found to
be useful for many other purpose such as collective foraging~\cite{stocker}.)
We also use three different values of the self-propelled 
velocity: $v_0 = 0$ (the passive case), $v_0 = 0.2u_\eta$, and $v_0 =
1.7u_\eta$, where $u_{\eta}$ is the characteristic small-scale (Kolmogorov) fluid velocity. Furthermore we use several values of
$\alpha$ and $St$ (which corresponds to particles, or microorganisms with sizes which vary between $1 \mu m$ to $50\mu m$) 
and perform simulations with $N_p = 50,000$ particles for every $\alpha, St$, $r_{\rm int}$, and $v_0$.

It is useful to begin with an intuitive picture of whether there is any evidence of
collective motion in our model. This is best seen online in videos~\cite{video} of the time
evolution of a collection of such microorganisms with different shapes, sizes,
activity ($v_0$) and ranges of interaction ($r_{\rm int}$) with random initial
positions.  These simulations show convincingly how flocking emerges only
when conditions are \textit{right} in terms of not only shapes and sizes but
also $v_0$ and $r_{\rm int}$. Indeed, for the cases where they do flock, a
snapshot of the velocity vectors of the individuals (Fig.~\ref{fig:snapshot})
clearly reflect the level of ordering in our model. For clarity, in
Fig.~\ref{fig:snapshot} we only show a randomly chosen subset of the total
$N_p$ rod-like ($\alpha = 2.0$) and inertial ($St = 2.0$) microorganisms 
in the flow.

With this picture in mind, and visual evidence of flocking (depending on the {\it type} of microorganism), we cannot 
refrain anymore from making our study quantitative. A convenient way to quantify the degree of flocking, is to measure the global order parameter~\cite{vicsek} 
\begin{equation}
\Phi_v=\left \langle \frac{1}{N_p}\sum_{j=1}^{N_p} {\bf{\hat{v}}}_j\right \rangle
\end{equation}
where the angular brackets denote a time-average (after the initial transients in their motion 
have died down) and ${\bf{\hat{v}}}_j$ is the unit velocity vector of the $j$-th individual. This 
order parameter ought to vary from species to species
of different shapes and sizes as well as depend on the level of activity and 
radius of interaction. Should organisms flock, then all velocity vectors 
must point in the same direction yielding $\phi_v = 1.0$. On the other hand, if there is no ordering, then the
velocity vectors ought to point in different directions for different individuals leading to $\Phi_v = 0.0$. Hence, since by definition 
$0.0 \le \Phi_v \le 1.0$, such a global order parameter quantifies the
degree of flocking in a species unambiguously. 

Figure~\ref{fig:global} shows representative pseudo-color
plots of $\Phi_v$ as a function of $\alpha$ and $St$ (on a log scale)
for different $r_{\rm int}$ and $v_0$ (see figure caption). 
These results are intriguing for several reasons. For individuals which are not
active, i.e., $v_0 = 0$, microorganisms flock only when they are sufficiently
large, rod-like and with a very large radius of interaction. This clearly shows
that in the absence of activity, it is extremely hard for microorganisms which
are small, slender or disk-shaped, to overcome turbulent mixing and drag to
self-organise. In this parameter space, even if an individual aligns with its
    neighbour, eventually that direction gets randomised due to the
    turbulent medium and therefore it cannot continue with the direction over time which
    is necessary for the flocking to emerge.   

However as soon as we turn on \textit{activity}, i.e., $v_0
\neq 0$ but still small (Fig.~\ref{fig:global}b), we obtain non-zero
values of the order parameter for all species only for large sizes ($St > 
1$). For smaller-sized microorganisms for a given radius of interaction, the
shape plays a crucial role: Rods are more likely to show collective behaviour
than disks because, as has been known, correlation time-scales for rods are 
considerably longer than disks~\cite{AmalPRE}.
Understandably, as the $r_{\rm int}$ and $v_0$ increase
(e.g., Fig.~\ref{fig:global}c), the degree
of flocking enhances for all species. Nevertheless, it is clear from our
results that both size and shape matter---in a non-monotonic and complicated
way---decisively in determining the ability of a species to overcome turbulent
mixing and drag to swarm. Indeed for 
a significant range of values of $\alpha$, $St$, $v_0$, and 
$r_{\rm int}$, we do find evidence of near perfect flocking ($\Phi_v \to 1.0$).

\begin{figure}
\begin{center}
\includegraphics[width=1.0\columnwidth]{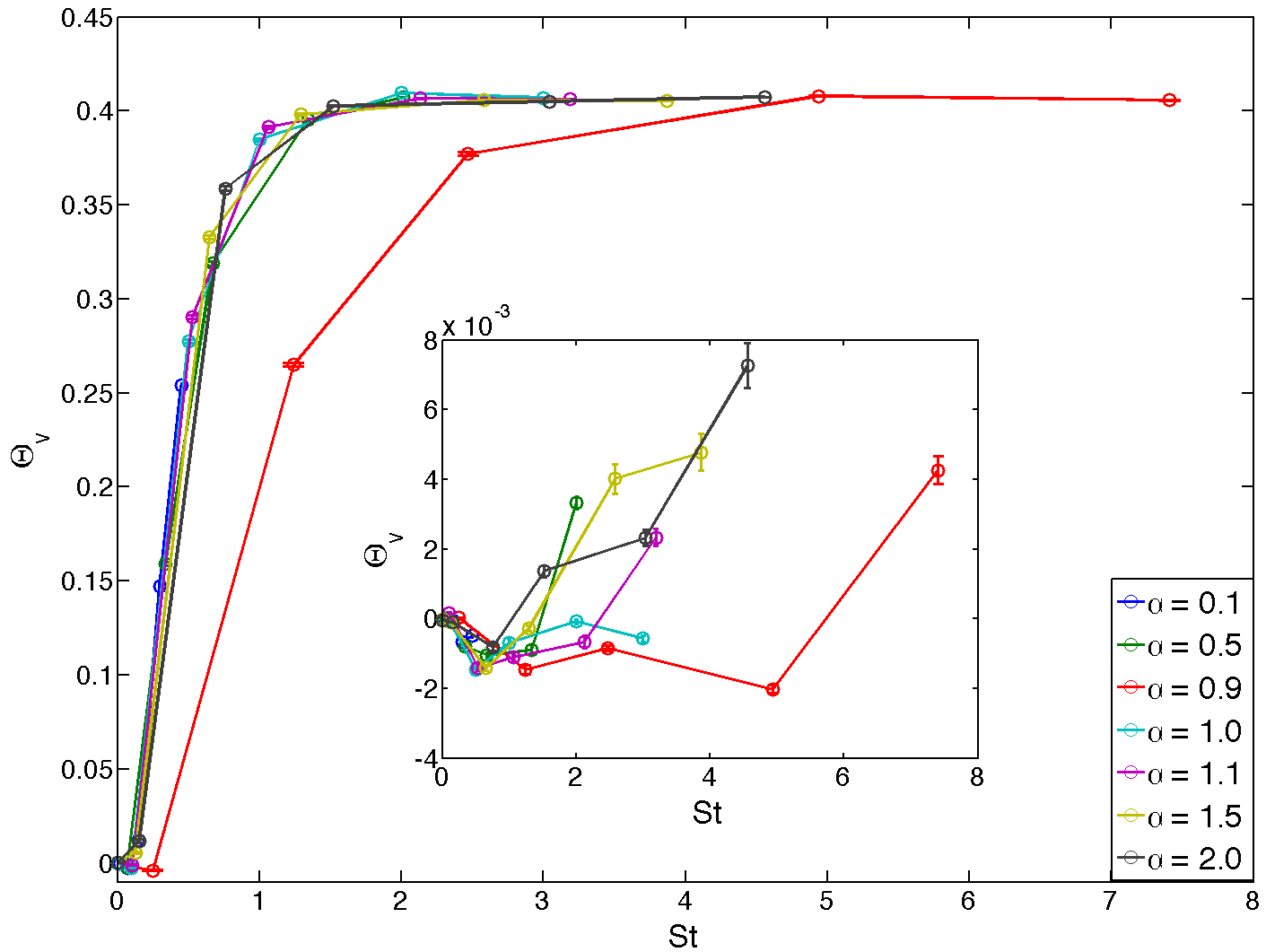}
\end{center}
\caption{Representative plot of the nematic order, as a function of size $St$, 
for differently shaped microorganisms $\alpha$ for $r_{\rm int} = 0.2\pi$. We show results for both active ($v_0 = 0.2u_\eta$) as well as (inset) passive
($v_0 = 0$) microorganisms.}
\label{fig:nematic}
\end{figure}

This is the first clear evidence of how collective behaviour can emerge in a
collection of microorganism in a complex, random turbulent medium. Before we
provide a theoretical explanation of this emergent phenomenon, it is useful to
characterise one additional aspect, namely the nematic order~\cite{chate}, of these flocks.
This is done most conveniently through the tensorial order parameter
\begin{equation}
{\bf{Q}}=\left \langle \hat{\bf{v}}_i\otimes \hat{\bf{v}}_j-\frac{1}{3}\bf{1}\right \rangle;
\end{equation} 
whence we can calculate the scalar order parameter, associated with the nematic
order, $S_v({\bf x})=n^v_{\alpha}n^v_{\beta}Q_{\alpha\beta}$,  where ${\bf
n}^{v}({\bf x})$ are the local symmetry directions of the fields.  Since we
are interested in the global behaviour of this system, we calculate the global 
nematic order $\Theta_v = \int d{\bf x}S_v$. In Fig.~\ref{fig:nematic} we show a representative plot of 
$\Theta_v$, for a given radius of interaction and self-propelled 
velocity as well as for, in the inset, the same $r_{\rm int}$ but for $v_0 = 0$, as a function of the Stokes number and for different 
shapes $\alpha$. Although we have not obtained perfect nematic ordering, we do see that a certain level of 
order emerges only for Stokes numbers larger than 1. From the inset, it is clear there is no ordering 
when particles are essentially passive but still allowed to interact over the same length scale. This is of course 
consistent with the picture that emerged when we looked at the global
order parameter $\Phi_v$.
(A similar definition is possible for the orientation vector ${\bf \hat p}$; we have checked that the fluctuations are 
far stronger for the orientation vector as discussed at the end of this paper.)

We know that finite-sized, passive, inertial particles tend to preferentially concentrate
for a range of Stokes numbers.  A convenient quantification of this
inhomogeneous distribution of particles in a flow is through the correlation
dimension $D_2$ which is measured via the probability of finding two particles
within a distance $r$, namely, $P^<(r) \sim r^{D_2}$.  For tracers ($St = 0$), which are
space filling, $D_2 = 3$ (in 3 dimensional flows). However for non-tracers
($St \neq 0$), it
decreases from $D_2 = 3$ with increasing $St$ and reaches a minima
for $St = \mathcal{O}(1)$, before increasing again to saturate at  $D_2 = 3$
for $St \gg 1$~\cite{bec2}.  It is this non-monotonic
(as a function of the Stokes number) nature of preferential concentration of
heavy, inertial particles which we show is central to flocking of our active particles.
Hence, the particle dynamics ensure that for certain values of $\alpha$
and $St$ more organisms are \textit{forced}, mechanically, to be close to each
other and hence an enhancement in the degree of flocking. Therefore if the microorganisms 
are active, such preferential concentration plays a dominant role in the emergence of 
flocking. In the absence of preferential concentration, the chances of a large fraction of microorganisms clustered close enough for 
their mutual alignment-interactions to become effective would be minimal.
 
Our findings could be of importance for fabrication and design of artificial
micro-swimmers. But are they equally relevant for observations in nature? 
Typical marine environments are known to be turbulent with
a Reynolds number comparable to the ones we use in our simulations and a
Kolmogorov scale $\eta$ of the order of a few $mm$~\cite{stocker,jimenez}. In such an environment, 
microorganisms---not necessarily spherical---such as
zooplankton often show collective behaviour~\cite{yen}. Recent measurements suggest that
this class of organisms have a variety of sizes which could range from 1$\mu mm$
to $10mm$ and, hence, less than $\eta$~\cite{stocker,yen}. This scale separation ensures that our
modeling of microorganisms through a combination of the linear Stokes drag
model~\cite{bec1,bec2,ray-PoF}, for the translational degrees of freedom and the Jeffery
equation~\cite{jeffery,AmalPRE,PumirNJP,ToschiPRL}, for
the rotational degrees of freedom, is valid. Additionally, the associated
response time scale, the Stokes time, $1\mu s \le \tau_p \le 1s$, of such
microorganisms when compared to the Kolmogorov time scale $\tau_\eta$ of the
ocean environment leads to Stokes numbers of the same order as we have used in our
study~\cite{yen}.

These arguments suggest that our theoretical framework, even with its
limitations (see below), is a relevant model for small organisms in a marine
environment. Our results show that
collective, ordered motion in active systems is an emergent
phenomenon which can spontaneously occur even in a
hostile environment where the range of interactions (here,
quantified by $r_{\rm int}$) for a collection of individuals is restricted.
These results show that purely mechanical principles related to the dynamics of
finite-sized particles in turbulence are critical in {\it forcing} certain
organisms to flock: Given the right size and shape, microorganisms are brought
in much closer proximity to each other allowing them
to behave cooperatively and flock. Indeed recent studies of passive, non-interacting, non-spherical
particles shed light on the correlation between particle trajectories and flow
directions~\cite{PumirNJP,ToschiPRL,AmalPRE}; as particles become interacting
and self-propelled the competing effects between the passive and active cases
lead to our very interesting---and surprising---discovery that shape and size
both matters for stabilising model flocks in turbulent flows. Without these
underlying principles of fluid mechanics a random, chaotic environment would
have pushed individuals far apart leading to a break up  of flocks. Hence, by
bringing together basic principles of turbulent transport and active matter, we
have, for the first time, shown how model flocks can form in a collection of
self-propelled individuals moving in a turbulent flow. Such cooperative behaviour 
has recently been reported for dry, granular systems~\cite{sood-nature} but 
not for the complex system that we report here. More pertinently, Durham, {\it et al.}~\cite{durham}, 
considered modelled active tracers to understand the observed \textit{patchiness} of microorganisms such 
as phytoplankton: However, the idea of collective motion and the role played by inertia as well as the translational 
and rotational degrees of freedom was largely ignored in this and other studies.

Before we conclude, it is essential to understand some of the limitations of
our model. Our model cannot capture the more dramatic instances of flocking in
nature which involve macroscopically large organisms such as birds. This class
of phenomena should be studied with a similar fluid mechanics approach in
future by resolving boundaries  and developing ideas for large structure-fluid
interactions. Furthermore, the role of hydrodynamic interactions in flock stability ought to be investigated
systematically within
the present model. Finally, a more systematic study of finite (population) size
effects and the effect of intrinsic noise as well as the level of activity is
left for future work.  

We hope that this novel---yet simple---mechanical approach, bringing together already well-established concepts in 
different areas,  is an important ingredient in explaining why flocking is
ubiquitous, and is as much a strategy as it is {\it forced}. Such an approach could also lead to new
ideas beyond the present study such as developing more realistic
predator-prey models in complex environment.

\begin{acknowledgements}
SSR acknowledges the support of the DAE, Indo--French Center for Applied
Mathematics (IFCAM) and the Airbus Group Corporate Foundation Chair in
Mathematics of Complex Systems established in ICTS.  AR and SSR acknowledges
the support of the DST (India) project ECR/2015/000361. The simulations were
performed on the cluster {\it Mowgli} and workstations {\it Goopy} and {\it
Bagha} at the ICTS-TIFR. AS acknowledges start-up grant (No.4-5
(206-FRP)/2015(BSR)) from UGC. 
\end{acknowledgements}

\end{document}